\documentstyle[12pt,bezier]{article}
\newcommand{\be}{\begin{equation}}
\newcommand{\ee}{\end{equation}}
\newcommand{\ba}{\begin{eqnarray}}
\newcommand{\ea}{\end{eqnarray}}
\newcommand{\ol}{\overline}
\newcommand{\ra}{\rightarrow}
\newcommand{\lb}[1]{\label{#1}}
\newcommand{\bb}[1]{\bibitem{#1}}

\begin{document}

\begin{titlepage}
\setcounter{page}{1}
\title{Flat wormholes from straight cosmic strings}
\author{G\'erard Cl\'ement\thanks{E--mail:
 gecl@ccr.jussieu.fr} \\
\small Laboratoire de Gravitation et Cosmologie Relativistes
 \\
\small Universit\'e Pierre et Marie Curie, CNRS/URA769 \\
\small Tour 22--12, Bo\^{\i}te 142,
 4 place Jussieu, 75252 Paris cedex 05, France}
\bigskip
\date{\small 28 January 1996}
\maketitle
\begin{abstract}
We describe the analytical extension of certain cylindrical multi--cosmic
string metrics to wormhole spacetimes with only one region at spatial
infinity, and investigate in detail the geometry of asymptotically Minkowskian
wormhole spacetimes generated by one or two cosmic strings. We find that
such wormholes tend to lengthen rather than shorten space travel. Possible
signatures of these wormholes are briefly discussed.
\end{abstract}
\end{titlepage}

\section{Introduction}

It has long been recognized that the equations of general relativity carry
information not only about the local geometry of spacetime, but also about
its possible global topologies. Early work on wormholes \cite{1}--\cite{4} was 
motivated
by the hope that they might provide a way to evade the problems associated
with point singularities in particle physics. A quite different motivation
was provided by the analysis of Morris and Thorne, who first investigated in
some detail the possibility of using traversable wormholes to travel in
space \cite{12}, as well as in time \cite{5}. 

Traversable wormholes may occur as solutions to the Einstein field equations
with suitable sources violating the weak energy condition. When explicit
solutions are discussed in the literature, these are usually static
spherically symmetric Einstein--Rosen wormholes connecting two regions at
spacelike infinity \cite{6}. More relevant for the purpose of interstellar
travel are Wheeler--Misner wormholes \cite{2}, with only one region at
spatial infinity. Exact solutions of the time--symmetric initial--value
problem of general relativity with such a topology have been constructed
\cite{3}\cite{4}, but these non--static wormholes are not traversable 
\cite{12}. In a recent paper \cite{MP},
Schein and Aichelburg have constructed a static Wheeler--Misner wormhole by
matching, along two spherical shells $S_1$ and $S_2$, an outer
Majumdar--Papapetrou spacetime to an inner Reissner--Nordstr\"{o}m
spacetime; this is traversable only one way, from $S_1$ to $S_2$.

Static Wheeler--Misner wormholes may be obtained by suitably extending a
procedure, described by Visser, to construct models of flat Einstein--Rosen
wormholes \cite{7}. Remove from Euclidean space $R^3$ a volume 
$ \Omega $ .\ Take a second, identical copy of $R^3-\Omega $, and
identify these two excised spaces along the boundaries $\partial \Omega $.
The spacetime obtained by factoring the resulting space with the time axis 
$ R $ is a geodesically complete Einstein--Rosen wormhole (or
multi--wormhole if $\Omega $ has several connected components), everywhere
flat except on $\partial \Omega $, where the stress--energy is concentrated.
To similarly construct a Wheeler--Misner wormhole, remove from $R^3$
two non--overlapping volumes $\Omega $ and $\Omega ^{\prime }$ which are the
image of each other under the involution $(x,y,z)\rightarrow (-x,y,z)$, and
identify the boundaries $\partial \Omega $, $\partial \Omega ^{\prime }$
(Fig.\  1) (the diffeomorphism $\Omega \rightarrow \Omega ^{\prime }$ must
reverse orientation if the resulting manifold is to be orientable \cite{8}).
In a further extension of this procedure, the boundaries $\partial
\Omega$, $\partial \Omega^{\prime}$ are not identified, but connected by a
cylindrical tube carrying equal energy per unit length and longitudinal
tension (if the surface $\partial \Omega$ is compact and simply connected,
it follows from the Gauss--Bonnet theorem that the energy per unit tube
length is $1/2G$). The geometry, as viewed by an ``external'' observer (in
$R^3 - \Omega - \Omega^{\prime}$), does not depend on the ``internal''
distance (through the tube) between the two ``mouths'' $\partial
\Omega$, $\partial \Omega^{\prime}$, which may be arbitrarily large, so
that the advantage for space travel is not so obvious.

In the case where $\partial \Omega$ is a cylinder, the internal tube is,
as well as external space, flat. The curvature is then concentrated on the
two wormhole mouths, each of which carries (again by the Gauss--Bonnet
theorem) a mass per unit length and a longitudinal tension both equal to
$m = -1/4G$. For instance, the static conical Einstein--Rosen wormholes
generated by a circular cylindrical source \cite{9} may thus, in the case of a
vanishing deficit angle, be extended to Wheeler--Misner wormholes with zero
tube length, which may easily be generalized to the case of an arbitrary
tube length.

Let us now discuss the case where $\Omega$ is a polyhedron. Visser showed
\cite{4} that in this case the curvature of the boundary $\partial \Omega$
is concentrated on the edges, which each carry an equal energy per unit
length and tension. A particular case of Visser's polyhedral wormholes is
obtained in the limit of cylindrical polyhedra, i.\ e.\  configurations of
$p$ parallel cosmic strings of tension $m_i$, with $\sum_i m_i = -1/2G$.
Following the procedure described above, these Einstein--Rosen wormholes
can be straigthforwardly be extended to Wheeler--Misner wormholes generated
by $p$ straight cosmic strings in the case of zero tube length, or $2p$
cosmic strings for an arbitrary tube length. 

The purpose of this paper is to investigate in more detail the
construction and geometry of flat cylindrical wormholes generated by
straight cosmic strings, following an analytical method complementary to
the geometrical method outlined above. In the second section we show how
special multi--cosmic string metrics may be analytically extended 
\cite{10}\cite{11} to Einstein--Rosen or Wheeler--Misner multi--wormhole,
multi--cosmic string metrics. Because our spacetime is locally
approximately Minkowskian, we shall be specially interested in
asymptotically Minkowskian spacetimes. In the third section, we study in 
detail the
topology and geometry of asymptotically Minkowskian, flat Wheeler--Misner
wormholes generated by one or two straight cosmic strings. Geodesic paths
through such wormholes are discussed in the fourth section, with
applications to space travel and geometrical optics. Our results are 
summarized and discussed in the last section.
 
\setcounter{equation}{0}

\section{Wormholes from cosmic strings}

We start from the well--known multi--cosmic string metric \cite{13}--\cite{15}
\be \lb{2.1}
ds^2 = dt^2 - d\sigma^2 - dz^2,
\ee
where the 2--metric
\be \lb{2.2}
d\sigma^2 = \prod_i |\zeta - a_i|^{-8Gm_i} \,d\zeta \,d\zeta^\ast
\ee
($\zeta \equiv x + iy$) may locally be transformed to the Cartesian form 
\be \lb{2.3}
d\sigma^2 = dw \,dw^\ast = du^2 + dv^2
\ee
($w \equiv u + iv$) with
\be \lb{2.4}
dw = \prod_i(\zeta - a_i)^{-4Gm_i} \,d\zeta.
\ee
The spacetime of metric (\ref{2.1}) is therefore flat outside the
worldsheets of the cosmic strings $\zeta = a_i$ (the conical
singularities, with deficit angle $8\pi Gm_i$, of the surface (\ref{2.2})),
which carry an energy per unit length and a tension both equal to $m_i <
1/4G$ (for $m_i \geq 1/4G$ the singularity $\zeta = a_i$ is at spatial
infinity). The metric (\ref{2.2}) is also generically singular at the
point at infinity in the complex $\zeta$--plane, with deficit angle $8\pi 
G\sum_i m_i$. This last singularity is at infinite geodesic distance if
\be \lb{2.5}
\sum_i m_i \leq 1/4G\,.
\ee

In the generic case, the conical singularities of the metric (2.2) are
logarithmic branch points which become branch points of order $n$ for
$4Gm_i = 1/n$ ($n$ integer). The conformal factor in (\ref{2.2}) is
analytical in the complex $\zeta$--plane with cuts extending from the
various branch points to infinity. Consider now the special case of the
bicone with $m_1 = m_2 = 1/8G$,
\be \lb{2.6}
d\sigma^2 = \frac{d\zeta \,d\zeta^\ast}{|\zeta^2 - b^2|}.
\ee
By choosing the cut to be the segment connecting the two branch points
$\zeta = \pm b$, we can analytically extend this bicone to a geodesically
complete surface: a cylinder. To show this \cite{11}, pinch the cylinder
along a parallel. We thus obtain two identical bicones with deficit angles
$\pi$ at the two vertices and $2 \pi$ at infinity, joined along the pinch. 
These two bicones are diffeomorphic to the two sheets of the Riemann
surface of the metric (\ref{2.6}) with the cut indicated (Fig.\  2). The
diffeomorphism is implemented by the transformation
\be \lb{2.7}
\zeta = b \cosh w
\ee
(by integration of (\ref{2.4})), which maps the cut into the circle $u = 0$
($v$ is an angular variable from (2.7)), and the two sheets into the two
halves $u > 0$ and $u < 0$ of the cylinder.

The cylinder with its two circles at infinity is the basic building block
for Einstein--Rosen wormholes in two space dimensions. The general
one--wormhole flat metric is obtained by multiplying the right--hand side of
(2.6) by an arbitrary conformal factor assumed to be regular at $\zeta =
\pm b$. In the simplest case,
\be \lb{2.8}
d\sigma^2 = \frac{|\zeta - c|^{-8Gm}}{|\zeta^2 - b^2|} \,d\zeta \,d\zeta^\ast,
\ee
take the cut to be the geodesic segment connecting the two branch points
$\zeta = \pm b$; analytical continuation across this cut then leads to a
surface with two symmetrical asymptotically conical sheets smoothly
connected along a cylindrical throat, and two conical singularities ---one
in each sheet--- at $\zeta = c$. The corresponding spacetime (\ref{2.1})
is therefore a two--cosmic string Einstein--Rosen wormhole. Note that the
masses per unit length $m$ of the sources are different from the ``total''
masses per unit length $M$ determined from the asymptotic behaviour of the
metric at either region at spatial infinity \cite{16}\cite{11},  
\be \lb{2.9}
M = \frac{1}{4G} + m,
\ee
the difference $1/4G$ being the topological contribution of the wormhole.
For the spatial sections to be open ($M \leq 1/4G$), $m$ must be negative
or zero.

In the special case $m = -1/4G$,
\be \lb{2.10}
d\sigma^2 = \frac{|\zeta-c|^2}{|\zeta^2 - b^2|} \,d\zeta \,d\zeta^\ast,
\ee
the cosmic string and wormhole contributions to the total mass balance
(for this reason we shall refer to the metric (\ref{2.10}) as the
``dipole'' metric), so that the flat metric (\ref{2.1}) is asymptotically
Minkowskian. In Fig.\  3 we show schematically the pattern of geodesics 
$u =$ const. and $v =$ const. for
the dipole geometry (\ref{2.10}) (in the case $b$ and $c$ real, $|b| <
|c|$). The critical geodesic $u = 0$ which hits the singularity at 
$\zeta = c$ divides
the $\zeta$--plane in three regions. The geometry of the ``left'' region
($u < 0$) is that of a half--plane. In the ``inner'' region ($u > 0$),
the geodesic cut $u = L$ is surrounded by concentric closed geodesics 
$u =$ const. with 
equal perimeter $2\pi c$, until the geodesic segment $u = 0$ connecting
the conical singularity with itself is reached;
this region is the map of a truncated cylinder. Finally, the
``right'' region ($u > 0$) is again the map of a half--plane, however the
geodesic distance between two distant geodesics $v = D$ and $v = -D$ is
smaller in the ``right'' region than in the ``left'' region by a length
$2 l$, with $l = \pi c$. 

We thus arrive at the following geometrical construction for the
$t =$ const., $z =$ const. sections of the dipole wormhole:
1) Remove from the ($u$, $v$) plane the semi--infinite strip $u > 0$, $-l
< v < l$, and glue together the edges ($u > 0$, $v = l$) and ($u
> 0$, $v = -l)$. This yields a flat surface which is the union of the
``left'' and ``right'' regions discussed above, and has a closed boundary
geodesic $\Gamma$ of length $2l$ (the segment ($u = 0$, $-l \leq v
\leq l$) connecting the singularity ($u = 0$, $v = \pm l$) to
itself). 2) Take a second identical surface with boundary $\Gamma'$, and
glue together the two boundaries $\Gamma$, $\Gamma'$ to the two ends of a
truncated cylinder of perimeter $2l$ and length $2L$, with
\be \lb{2.11}
L = \sqrt{c^2 - b^2} - c\log[(c + \sqrt{c^2 - b^2})/b].
\ee
The resulting flat surface has the Einstein--Rosen wormhole topology, is
asymptotically Euclidean with two conical singularities, and is mapped on
the two--sheeted $\zeta$--plane by the analytical extension of (\ref{2.10})
described above. In the limit $L \rightarrow 0$ ($c \rightarrow b$), the
two singularities coalesce to a single conical singularity; this wormhole
can be viewed as a very special case of Visser's polyhedral wormholes, a
monohedron with one edge (the straight cosmic string) bounding one face
($\Gamma \times R$) connecting the two sheets.

The general $n$--wormhole metric is obtained by first making on the
cylinder metric (\ref{2.6}) the conformal transformation $\zeta
\rightarrow P_n(\zeta)$, where $P_n$ is a polynomial of order $n$, then  
multiplying the resulting $n$--cylinder metric by a conformal factor
regular at the zeroes of $P_n(\zeta) \mp b$. Multi--wormholes generated by
a sigma--model field coupled repulsively to gravity are discussed in
\cite{11}. The metric for a flat $n$--wormhole spacetime generated by $p$
cosmic strings is (\ref{2.1}) with
\be \lb{2.12}
d\sigma^2 = \frac{\prod_{i=1}^p |\zeta - c_i|^{-8Gm_i}}{|\zeta_n^2 - b^
{2n}|} \,d\zeta \,d\zeta^\ast,
\ee
where
\be \lb{2.13}
\zeta_n = \prod_{j=1}^n (\zeta - a_j).
\ee
Various extensions of this metric across the $n$ cuts are possible. In the
symmetrical extension, the Riemann surface is made of two sheets joined
along the $n$--component cut. The resulting spacetime is an $n$--wormhole
Einstein--Rosen spacetime, with $p$ cosmic strings in each sheet, and total 
mass per unit length
\be \lb{2.14}
M = \frac{n}{4G} + \sum_{i=1}^p
m_i,
\ee
in accordance with the Gauss--Bonnet theorem \cite{11}.

The simplest case after $n = 1$ is $n = 2, p = 0$:
\be \lb{2.15}
d\sigma^2 = \frac{l^4 \, d\zeta \, d\zeta^\ast}{|(\zeta^2 - a^2)^2 - b^4|}\,. 
\ee
Then, $M = 1/2G$, so that the two--dimensional spatial sections of genus 1,
orientable by construction, are compact and regular, i.e.\ are tori $T_1 =
S^1 \times S^1$. To recover the symmetrical Riemann surface, 
pinch the torus along two opposite circles; this yields two tetracones
with deficit angles $\pi$ at each vertex, joined along the two pinches,
which correspond to the two cuts of the Riemann surface. The flat metric 
(\ref{2.3}) on the torus, with $u$ and $v$ periodical, is transformed to
the metric (\ref{2.15}) by
\be \lb{2.16}
\zeta (w) = \sqrt{a^2 + b^2} \, {\rm sn} \left( \frac{\sqrt{a^2+b^2}}{l^2}w, k
\right)
\ee
(with $k^2 = (a^2 - b^2)/(a^2 + b^2)$), where sn is a biperiodical Jacobi 
function.

However the metric (\ref{2.15}) admits a more economical analytical
extension to a topologically non trivial Riemann surface with only one
sheet. Such a possibility derives from the observation \cite{11} that the
torus may be pinched only once into a single tetracone joined to itself by
an identification of the two edges. This identification corresponds to an
identification of the two cuts, leading to the identification of the two
sheets, of the Riemann surface for the complex variable $\zeta(w)$, the
point $\zeta$ in the first sheet of the symmetrical extension being
identified with the point $-\zeta$ in the second sheet \footnote{Other
possible identifications  $\zeta \rightarrow \pm 
\zeta^{\ast}$ between the two sheets
would lead to the non--orientable manifold $U_2$.}. A large
circle geodesic $ v = $ const. is thus mapped into a line connecting the
two cuts either in the upper or in the lower half--plane of the
$\zeta$--plane, so that a particle going around the torus along this
geodesic falls into e.g. the left--hand cut to come out again from the
right--hand cut (Fig.\  4).

Such a one--sheeted extension is possible whenever the distribution of both
the $n$ cuts and the $p$ conical singularities of the flat metric (\ref{2.12})
is invariant under the isometry $\zeta \rightarrow - \zeta$, so that the
two sheets of the symmetrical extension may be identified together. In the
case $n = 2$ the resulting surface ---a topological torus with a point at
infinity (provided $M < 1/4G$) and $p$ conical singularities--- is a 
Wheeler--Misner wormhole. In
fact our construction is the three--dimensional counterpart of Lindquist's
\cite{4} reinterpretation of a four--dimensional Einstein--Rosen manifold
with two identical spherical bridges as a single Wheeler--Misner wormhole
by identifying corresponding points on the two sheets of the
Einstein--Rosen manifold . In the next section we investigate the various 
possible geometries for asymptotically Minkowskian ($M = 0$) Wheeler--Misner
wormholes with $p = 1$ or 2.

\setcounter{equation}{0}
\section{Two--string and one--string asymptotically Minkowskian
Wheeler--Misner wormholes}   
The Wheeler--Misner wormhole generated by two cosmic strings is the
one--sheeted extension of a symmetrical $n = p = 2$ metric (\ref{2.12}). In
the asymptotically Minkowskian case, this metric
\be \lb{3.1}
d\sigma^2 = \frac{|\zeta^2 - c^2|^2}{|(\zeta^2 - a^2)^2 -
b^4|}\,d\zeta\,d\zeta^{\ast} 
\ee 
depends on three complex parameters $a$, $b$, $c$. According to the
relative values of these parameters, the non--extended geometrical
configuration may belong to one of three possible generic types DD, AA,
or Q.

1) Dipole--dipole (DD). The sequences of closed ``u--geodesics'' surrounding
each of the two symmetrical geodesic cuts terminate in two disjoint
geodesic segments, each connecting one of the singularities with itself. A
single v--geodesic segment, bissecting the angles formed by the
continuations of these critical u--geodesics, connects the two
singularities together. An instance of this case is $a$, $b$, $c$ real,
$a^2 > b^2 + c^2$ (Fig.\  5(a)). The analytical extension of this geometry
to the Riemann surface obtained by identification of the two cuts, as
described in the previous section, leads to the DD wormhole geometry. The
geometrical construction of this wormhole (Fig.\  5(b)) follows closely that
of the dipole Einstein--Rosen wormhole, except that the two
copies of the $(u, v)$ plane deprived of a semi--infinite strip are replaced
by a single Euclidean plane deprived of a rectangular strip. Two opposite
edges, of length $2d$, of this rectangle are glued together, while the
other two edges, of length $2l$, are glued to the two ends of a truncated
cylinder of circumference $2l$ and length $2L$.

2) Antidipole--antidipole (AA). In this type again, two disjoint critical
u--geodesic segments connecting each of the two singularities with itself
enclose concentric closed u--geodesics surrounding a cut. However the two
singularities are now connected by two symmetrical v--geodesic segments,
bissecting the two angles formed by a critical closed u--geodesic segment
and its continuation to infinity (Fig.\  6(a), drawn for $a$, $b$, $c$ real,
$b^2 < a^2 < c^2 - b^2$). The corresponding Wheeler--Misner wormhole
geometry turns out to be equivalent to that of the Q wormhole, as we shall
presently explain.

3) Quadrupole (Q). In this case the sequences of closed u--geodesics 
surrounding the
two cuts terminate in two contiguous geodesic contours made from three
u--segments connecting the two singularities together (Fig.\  6(b), drawn for
$a$ and $b$ real with $b^2 < a^2$, and $c$ imaginary). The critical
v--geodesics bissecting the two angles formed by this self--intersecting
u--geodesic are $v = d$ and $v = - d$, where $2d$ is the
geodesic distance between the two singularities along the ``external''
segments of the u--geodesic $u = 0$. The distance between the two
singularities along the ``central'' segment $u = 0$ is $2(l - d)$, where
$l$ is again the perimeter of the closed u--geodesics. 
The geometrical construction of the Q Wheeler--Misner wormhole
resulting from identification of the two cuts is shown in Fig.\  7. The
Euclidean plane is incised  along a segment $AB$ of length $2d$. A torus
of ``small'' perimeter $2l$ ($l > d$) and ``large'' perimeter $2L$ is also
incised along a matching segment $A'B'$ (length $2d$) of a small circle.
Finally, the torus and the plane are glued together along the two edges of 
the cuts $AB$, $A'B'$. These two edges correspond to the two external
geodesic segments $u = 0$ in Fig.\  6(b); the complementary small--circle
segment connecting $A'$ and $B'$ is mapped into the central segment $u =
0$ in Fig.\  6(b), while the antipodal small circle on the torus is mapped
into the two identified cuts of the complex plane.

Now we show the equivalence of the Wheeler--Misner analytical extensions of
the AA and Q metrics of Fig.\  6, by showing that they correspond to two dual
maps for the same basic geometry in Fig.\  7. Instead of cutting the torus
along the antipodal small circle (which leads to the Q map of Fig.\  6(b)),
cut it along the large circle through O in Fig.\  7. In the Q map, this large
circle is, as all large circles, a v--geodesic connecting together the two
identified cuts of Fig.\  6(b). In the AA map, the same large circle now
corresponds to the two identified cuts of Fig.\  6(a), which are connected
together by a sequence of small circles --- now v--geodesics ---
terminating in the two sides $v = \pm l$ of the segment $A'B'$ ($AB$); the
two critical geodesics $v = \pm d$ of Fig.\  6(b) --- large circles going
through $A'$ and $B'$ --- correspond to the two critical geodesics $u =
\pm d$ of Fig.\  6(a). In other words, the AA and Q maps are transformed into
each other under the duality $u \leftrightarrow v$ exchanging the two
circles of $S^1 \times S^1$.

Asymptotically Minkowskian wormholes generated by a single cosmic string
may be obtained from the two--cosmic string case by taking limits such that
the two cosmic strings (the two singularities in the two--dimensional
spatial sections) coincide. Two inequivalent geometries may result, DD$_0$
or 8. Consider first the DD wormhole geometry, Fig.\  5(b), and take the
limit $L \ra 0$. The resulting ``DD$_0$'' geometry may be directly
obtained from the Euclidean plane deprived of a rectangular strip by
gluing together, first two opposite edges of the rectangle, then the other
two edges. Clearly, by construction the singularity is connected to itself
by only two geodesics of lengths $2d$ and $2l$. A metric which leads to
this geometry after a one--sheeted analytical extension is (\ref{3.1}) with
$a, b, c$ real, $a^2 = b^2 + c^2$.

Taking the limit $d \ra 0$ in the DD geometry of Fig.\  5(b) is obviously
equivalent to taking the limit $d \ra l$ in the Q geometry of Fig.\  7. The
resulting ``8'' geometry (Fig.\ 8) is obtained by incising the Euclidean 
plane along
a segment of length $2l$, bringing together the two vertices so that the
two edges make a figure 8, then gluing these two edges to the two ends of
a truncated cylinder of circumference $2l$ and length $2L$. The
singularity is connected to itself by $(2 + N)$ geodesics, two geodesics
of length $2l$ (the two edges just mentioned), and a denumerable family of
geodesics of length $2 \sqrt{L^2 + n^2 l^2}$ going from one end of the
cylinder to the other while winding $n$ times around it. Two examples of
metrics (\ref{3.1}) leading to this geometry are given by real parameters
$a, b, c$ with $b^2 < a^2$, and either $c^2 = 0$ or (because of the
equivalence AA $\leftrightarrow$ Q) $c^2 = a^2 + b^2$.

The two preceding one--cosmic string geometries depend on two parameters.
By taking the further limit $d = 0$ in the DD$_0$ geometry, or $L = 0$ in
the 8 geometry, we obtain the ``I'' geometry, which corresponds simply to
a plane with two points identified. A single geodesic, of length $2l$,
connects the resulting conical singularity to itself. This limiting
geometry may be obtained from the one--sheeted extension of the metric
\be
d\sigma^2 = \frac{|\zeta|^2}{|\zeta^2 - l^2|} \,d\zeta d\zeta^{\ast} \,.
\ee

\setcounter{equation}{0}
\section{Space travel and geometrical optics}

Because our wormhole spacetimes are (almost everywhere) flat, as well as
asymptotically Minkowskian, they do not classically scatter test particles
or light rays. A test particle going through a wormhole will emerge with a
worldline parallel to its ingoing worldline. However the outgoing
worldline will generically be shifted, in space as well as in time. Shifts
in space lead to non--trivial geometrical optics effects, while shifts in
time might be relevant for, e.g., intergalactic travel. We first consider
shifts in time, with a view to address the question, raised in the
Introduction, whether traversable wormholes are really advantageous for
long distance space travel. Because of the simple form of our spacetime
metric (\ref{2.1}), time shifts only depend on the three--velocity of the
test particle and on the geodesic distance travelled in two--dimensional
sections $t =$ const., $z =$ const. So we consider some given
two--dimensional geodesic as ``start'' line, and another, parallel geodesic
as ``finish'' line, and compare the geodesic distance between these two
lines along two paths, one which ``goes through the wormhole'', and
another which does not go through the wormhole.

First we have to give a workable definition of ``going through the
wormhole''. These questions are usually addressed in the context of
Einstein--Rosen wormholes, where a path which goes from one point at
spatial infinity to the other obviously ``goes through the wormhole''.
More generally, consider a space $E$ with $N$ points at spatial infinity. We
compactify this space to a closed topological space $\ol E$, and define a
``path going through a wormhole'' as a path, going from spatial infinity
to spatial infinity, which is not homotopic to zero. This definition
covers in particular both the case of the Einstein--Rosen wormhole, a
topological sphere with two points at infinity (paths connecting these two
points are trivially non contractible), and that of the Wheeler--Misner
wormholes of the previous section, topological tori with one point at
infinity (closed paths are non--contractible if they wind around one or
both circles).

For simplicity, we first deal with the case of one cosmic string. Fig.\ 9
shows two parallel geodesics 1 and 2 going through a
DD$_0$ wormhole of parameters ($l, d$); an effect of the shifts in space
mentioned above is that geodesic 1 comes in to the right of geodesic 2 but
comes out to its left. These geodesics cannot be deformed to the spectator
geodesics 3 or 4 without crossing the singularity. Both are shorter than
the spectator geodesics, the path being shorter by $2d \cos \theta$ for
geodesic 1, and by $2l \sin \theta$ for geodesic 2 ($\theta$ is the incidence
angle of geodesic 1). So we have here a model of a one--cosmic string
wormhole which does indeed shorten space travel. There is however a
hazard: a moving object, such as a spaceship, assumed to have a size of
the order of the dimensions $l$, $d$ of the wormhole, and to be light enough 
so as not to affect the geometry, would be cut in four pieces by the cosmic
string. 

Consider now the case of the 8 wormhole (Fig.\ 8). Obviously
a geodesic hitting, under the incidence $\theta$, one edge of the incision
in the Euclidean plane then goes the full length L of the cylinder, while
winding a number of times around it before emerging from the other,
contiguous edge. So travel through the wormhole will always be longer in
this case. We find that the path excess is, for a geodesic path, 
\be
\Delta = 2L \cos \theta + 2nl \sin \theta \,,
\ee
where $n = [(L/l) \tan \theta]$ is the winding number of the path, i.e.
the integer part of the number of turns
inside the cylinder. This number increases without 
limit as the incidence angle $\theta$ nears $\pi/2$, so that
\be
\Delta \simeq \frac{2L}{\cos \theta} \qquad {\rm for} \;\; \theta \ra \pi/2
\ee
becomes arbitrarily large. Finally, the I wormhole can be obtained as a limit 
of both the
DD$_0$ and the 8 wormholes so that, while technically a wormhole according
to our general definition, it is without effect on space travel.

Now for the two-cosmic string wormholes. In the DD geometry of Fig.\ 5(b) there 
are, as
in the DD$_0$ case, two kinds of paths through the wormhole, corresponding to
the two circles of $S^1 \times S^1$. Geodesic paths crossing once the two
identified segments of length $2d$ in Fig.\ \ 5(b) are ``shorter'' 
(than they would be in Euclidean space) by $2l\sin\theta$. Geodesic paths 
crossing once the two circular junctions of length $2l$ may wind $n$ times
around the cylinder; they can be shorter if $L < d$, but are always longer if
$L > d$, the path excess being given in terms of the winding number by
\be
\Delta = 2(L-d)\cos\theta + 2nl\sin\theta.
\ee

Paths through the Q wormhole of Fig.\ 7 are those which enter the torus
through one edge of the incision $AB$ and emerge through the other edge
after winding $N$ times around the large circle and $n$ times around the
small circle. The path excess for a geodesic path of incidence $\theta$ is
now
\be
\Delta = 2NL\cos\theta + 2nl\sin\theta\,,
\ee
with the relation
\be
n = [N(L/l)\tan\theta]
\ee
between the two winding numbers. The probability $p$ of a random geodesic
exiting the torus after one turn around the large circle is proportional
to the width of the gate $A'B'$, $p = d/l$, leading to the mean number of
turns around the large circle,
\be
\ol{N} = \frac{l}{d}\,.
\ee
So the path excess becomes very large in the limit of a very small
gatewidth ($d \ll l$) or of a grazing incidence ($\theta \simeq \pi/2$);
when both limits are taken, we obtain from Eqs.\ (4.4)--(4.6)
\be
\Delta \simeq \frac{2NL}{\cos\theta} \simeq \frac{2Ll}{d\cos\theta}\,.
\ee

Light rays from one geometrical point (e.\ g.\ a galactic source $S$) to
another (e.\ g.\ an observer $O$) may similarly follow a variety of
homotopically inequivalent optical paths (geodesics), leading to an array
of geometrical images $S'$. From Fig.\ 9, the DD$_0$ wormhole behaves
as a rectangular prism of infinite refractive index, and thus gives two images 
of a point source. The 8 wormhole behaves rather as a parallel plate with
partially reflecting faces, multiple reflections being 
replaced by multiple turns around the cylinder; the result is that a point
source gives rise to a one--dimensional periodical array of images. In the
case of the DD geometry, the observer would see a single ``near'' image
(due to light rays crossing the two identified segments of length $2d$ in
Fig.\ 5(b), equivalent to a parallel plate of infinite index) together
with a one--dimensional array of increasingly distant images (due to light
rays winding around the cylinder). Finally, in the case of the Q wormhole, 
the possibility of light rays winding
around both the large circle and the small circle of the torus leads to a
two--dimensional array of images $S'$, as shown in Fig.\ 10 (where 
the observer and source are assumed to be in the same plane $z = $ const.).

\setcounter{equation}{0}
\section{Conclusion}
We have shown that certain analytical maps may be extended to describe
Wheeler--Misner wormholes (with only one region at spatial infinity) which
are everywhere flat except for parallel cosmic string singularities.
We have used these analytically extended maps to investigate the
asymptotically Mink--owskian one--wormhole geometries generated by two 
cosmic strings
(these are either of the DD or Q type), or by a single cosmic string (of
the DD$_0$ or 8 type). As anticipated in the Introduction, it appears that 
such wormholes would have on the mean the
effect of lengthening rather than shortening space travel. 

Because of this
lengthening, which could be arbitrarily large, and of the unpleasant
consequences of accidentally hitting one of the cosmic strings, a space
traveller might wish to avoid these wormholes altogether. The presence of
such hypothetical wormholes as well as their type could in principle be  
inferred from the pattern of
images of a source viewed through the wormhole. The DD$_0$ wormhole (the
only one to always shorten space travel) would give only two images, while
other wormholes (which generically lengthen space travel) would give one-- 
or two--dimensional arrays of images. Such a characterization is
incomplete: most of the images could be too faint to be detected, or
hidden behind other objects. A wave--optics treatment should make possible
a better characterization. One anticipates
non--trivial effects arising both from diffraction by the topological
defects (cosmic strings) and resonance due to periodicity conditions in the
cylinder (or torus).

Our construction of static wormholes generated by staight cosmic strings
could be extended in two directions. A first extension should be to
construct wormholes generated by non--parallel moving straight cosmic
strings, along the lines of the analytical construction \cite{letgal} of
spacetimes generated by multiple moving crossed cosmic strings, and to
investigate the causal structure of these spacetimes.

Another possible extension would be to investigate wormholes generated by
closed cosmic strings or rings. As pointed out by Visser \cite{7}, the
polyhedon $\Omega$ mentioned in the Introduction can be collapsed to a
dihedron, or an irregular two--sided disk connecting two copies of $R^3$.
The resulting Einstein--Rosen wormhole is generated by the cosmic ring
circumscribing the disk. Some time ago, Zipoy \cite{zip} constructed
analytical static solutions to the vacuum Einstein equations with a
circular ring singularity and a double--sheeted topology. More recently,
Bronnikov and co--workers \cite{kb} have similarly constructed static
Einstein--Rosen wormhole solutions to $D$--dimensional gravity with a
circular ring singularity. However, the analytical construction of
Wheeler--Misner ring wormholes has not yet been attempted. Hopefully, this
problem could be addressed along the lines followed here for straight
cosmic strings.

%\end{document}

\newpage

%\documentstyle[bezier]{article}
%\textheight220mm\textwidth160mm\hoffset-15mm\voffset-25mm
%\begin{document}
\unitlength 1.00mm
\begin{picture}(80.00,80.00)(10,-25)
\put(20.00,25.00){\line(1,1){15.00}}
\put(65.00,25.00){\line(1,1){15.00}}
\put(20.00,8.00){\line(1,1){15.00}}
\put(65.00,8.00){\line(1,1){15.00}}
\put(35.00,40.00){\line(1,0){45.00}}
\put(20.00,25.00){\line(1,0){45.00}}
\put(35.00,23.00){\line(1,0){6.00}}
\put(59.00,23.00){\line(1,0){21.00}}
\bezier{16}(41.00,23.00)(50.00,23.00)(59.00,23.00)
\put(20.00,8.00){\line(1,0){45.00}}
\bezier{90}(41.00,32.00)(48.00,37.00)(59.00,35.00)
\bezier{60}(41.00,32.00)(44.00,28.00)(49.00,32.00)
\bezier{60}(49.00,32.00)(54.00,30.00)(59.00,35.00)
\bezier{15}(41.00,15.00)(48.00,20.00)(59.00,18.00)
\bezier{60}(41.00,15.00)(44.00,11.00)(49.00,15.00)
\bezier{60}(49.00,15.00)(54.00,13.00)(59.00,18.00)
\bezier{7}(41.00,32.00)(41.00,28.50)(41.00,25.00)
\put(41.00,25.00){\line(0,-1){10.00}}
\bezier{10}(59.00,35.00)(59.00,30.00)(59.00,25.00)
\put(59.00,25.00){\line(0,-1){7.00}}
\bezier{10}(42.00,31.00)(43.30,32.30)(44.60,33.60)
\bezier{15}(43.90,30.50)(45.90,32.50)(47.90,34.50)
\bezier{15}(45.80,30.60)(48.00,32.80)(50.20,35.00)
\bezier{12}(49.10,32.10)(50.60,33.60)(52.10,35.10)
\bezier{12}(50.90,31.90)(52.50,33.50)(54.10,35.10)
\bezier{12}(52.90,31.90)(54.40,33.40)(55.90,34.90)
\bezier{6}(56.00,33.00)(56.90,33.90)(57.80,34.80)
\bezier{10}(42.00,14.00)(43.30,15.30)(44.60,16.60)
\bezier{15}(43.90,13.50)(45.90,15.50)(47.90,17.50)
\bezier{15}(45.80,13.60)(48.00,15.80)(50.20,18.00)
\bezier{12}(49.10,15.10)(50.60,16.60)(52.10,18.10)
\bezier{12}(50.90,14.90)(52.50,16.50)(54.10,18.10)
\bezier{12}(52.90,14.90)(54.40,16.40)(55.90,17.90)
\bezier{6}(56.00,16.00)(56.90,16.90)(57.80,17.80)
\put(5.00,32.50){\makebox(0,0)[lc]{\large $R^3-\Omega$}}
\put(5.00,15.50){\makebox(0,0)[lc]{\large $R^3-\Omega$}}
\put(65.00,35.00){\makebox(0,0)[lc]{\large $\partial\Omega$}}
\put(65.00,18.00){\makebox(0,0)[lc]{\large $\partial\Omega$}}
\put(20.00,32.50){\line(1,0){13.00}}
\put(20.00,15.50){\line(1,0){13.00}}
\put(60.00,35.00){\line(1,0){4.00}}
\put(60.00,18.00){\line(1,0){4.00}}
\put(45.20,2.00){\makebox(0,0)[cc]{(a)}}
\put(113.00,20.00){\line(1,1){15.00}}
\put(155.00,25.00){\line(1,1){15.00}}
\put(155.00,8.00){\line(1,1){15.00}}
\put(138.00,40.00){\line(1,0){32.00}}
\put(123.00,25.00){\line(1,0){32.00}}
\put(149.00,23.00){\line(1,0){21.00}}
\bezier{12}(138.00,23.00)(143.50,23.00)(149.00,23.00)
\put(123.00,8.00){\line(1,0){32.00}}
\bezier{50}(128.00,35.00)(133.00,39.00)(138.00,40.00)
\bezier{50}(113.00,20.00)(118.00,24.00)(123.00,25.00)
\bezier{10}(114.00,10.00)(115.00,9.40)(115.80,9.00)
\bezier{30}(116.00,9.00)(119.50,8.35)(123.00,8.00)
\bezier{10}(129.00,25.00)(130.00,24.40)(130.80,24.00)
\bezier{7}(131.00,24.00)(134.50,23.35)(138.00,23.00)
\bezier{60}(113.00,20.00)(109.00,15.00)(114.00,10.00)
\bezier{90}(131.00,32.00)(138.00,37.00)(149.00,35.00)
\bezier{60}(131.00,32.00)(134.00,28.00)(139.00,32.00)
\bezier{60}(139.00,32.00)(144.00,30.00)(149.00,35.00)
\bezier{15}(131.00,15.00)(138.00,20.00)(149.00,18.00)
\bezier{60}(131.00,15.00)(134.00,11.00)(139.00,15.00)
\bezier{60}(139.00,15.00)(144.00,13.00)(149.00,18.00)
\bezier{7}(131.00,32.00)(131.00,28.50)(131.00,25.00)
\put(131.00,25.00){\line(0,-1){10.00}}
\bezier{10}(149.00,35.00)(149.00,30.00)(149.00,25.00)
\put(149.00,25.00){\line(0,-1){7.00}}
\bezier{10}(132.00,31.00)(133.30,32.30)(134.60,33.60)
\bezier{15}(133.90,30.50)(135.90,32.50)(137.90,34.50)
\bezier{15}(135.80,30.60)(138.00,32.80)(140.20,35.00)
\bezier{12}(139.10,32.10)(140.60,33.60)(142.10,35.10)
\bezier{12}(140.90,31.90)(142.50,33.50)(144.10,35.10)
\bezier{12}(142.90,31.90)(144.40,33.40)(145.90,34.90)
\bezier{6}(146.00,33.00)(146.90,33.90)(147.80,34.80)
\bezier{10}(132.00,14.00)(133.30,15.30)(134.60,16.60)
\bezier{15}(133.90,13.50)(135.90,15.50)(137.90,17.50)
\bezier{15}(135.80,13.60)(138.00,15.80)(140.20,18.00)
\bezier{12}(139.10,15.10)(140.60,16.60)(142.10,18.10)
\bezier{12}(140.90,14.90)(142.50,16.50)(144.10,18.10)
\bezier{12}(142.90,14.90)(144.40,16.40)(145.90,17.90)
\bezier{6}(146.00,16.00)(146.90,16.90)(147.80,17.80)
\put(84.00,15.50){\makebox(0,0)[lc]{\large $R^3-\Omega-\Omega'$}}
\put(155.00,35.00){\makebox(0,0)[lc]{\large $\partial\Omega$}}
\put(155.00,18.00){\makebox(0,0)[lc]{\large $\partial\Omega'$}}
\put(107.00,15.50){\line(1,0){8.00}}
\put(150.00,35.00){\line(1,0){4.00}}
\put(150.00,18.00){\line(1,0){4.00}}
\put(135.20,2.00){\makebox(0,0)[cc]{(b)}}
\end{picture}

\noindent
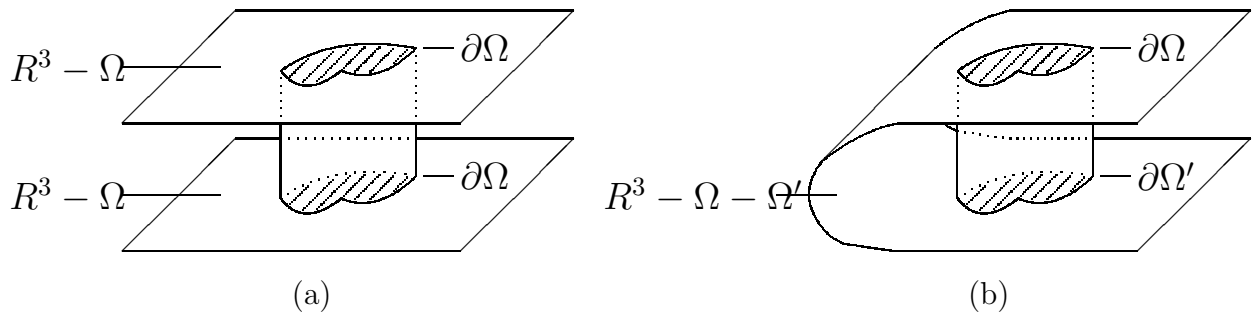
{\bf Fig.\ 1:} Construction of almost everywhere flat wormholes: (a)
Einstein--Rosen wormholes; (b) Wheeler--Misner wormholes.   
%\end{document}

%\documentstyle[bezier]{article}
%\textheight220mm\textwidth160mm\hoffset-15mm\voffset-25mm
%\begin{document}
%\unitlength 1.00mm
\begin{picture}(180.00,80.00)(10,-5)
\put(5.00,40.00){\line(0,-1){30.00}}
\put(25.00,40.00){\line(0,-1){30.00}}
\bezier{60}(9.00,42.00)(15.00,43.60)(21.00,42.00)
\bezier{60}(9.00,38.00)(15.00,36.40)(21.00,38.00)
\bezier{50}(9.00,42.00)(1.30,40.00)(9.00,38.00)
\bezier{50}(21.00,42.00)(28.70,40.00)(21.00,38.00)
\bezier{60}(9.00,8.00)(15.00,6.40)(21.00,8.00)
\bezier{15}(5.10,10.00)(7.00,8.50)(9.00,8.00)
\bezier{15}(24.90,10.00)(23.00,8.50)(21.00,8.00)
\bezier{15}(9.00,12.00)(15.00,13.60)(21.00,12.00)
\bezier{5}(5.10,10.00)(7.00,11.50)(9.00,12.00)
\bezier{5}(24.90,10.00)(23.00,11.50)(21.00,12.00)
\bezier{60}(9.00,23.00)(15.00,21.40)(21.00,23.00)
\bezier{15}(5.10,25.00)(7.00,23.50)(9.00,23.00)
\bezier{15}(24.90,25.00)(23.00,23.50)(21.00,23.00)
\bezier{15}(9.00,27.00)(15.00,28.60)(21.00,27.00)
\bezier{5}(5.10,25.00)(7.00,26.50)(9.00,27.00)
\bezier{5}(24.90,25.00)(23.00,26.50)(21.00,27.00)
\bezier{15}(5.10,36.00)(7.00,34.50)(9.00,34.00)
\bezier{60}(9.00,34.00)(15.00,32.40)(20.60,33.80)
\bezier{15}(24.90,36.00)(23.00,34.50)(21.10,34.05)
\put(9.00,38.00){\line(0,-1){24.50}}
\put(9.00,12.90){\line(0,-1){4.90}}
\put(31.00,28.00){\line(1,0){13.50}}
\put(38.00,22.00){\makebox(0,0)[c,c]{pinch}}
\put(50.00,40.00){\line(0,-1){30.00}}
\put(70.00,40.00){\line(0,-1){30.00}}
\bezier{60}(54.00,42.00)(60.00,43.60)(66.00,42.00)
\bezier{60}(54.00,38.00)(60.00,36.40)(66.00,38.00)
\bezier{50}(54.00,42.00)(46.30,40.00)(54.00,38.00)
\bezier{50}(66.00,42.00)(73.70,40.00)(66.00,38.00)
\bezier{60}(54.00,8.00)(60.00,6.40)(66.00,8.00)
\bezier{15}(50.10,10.00)(52.00,8.50)(54.00,8.00)
\bezier{15}(69.90,10.00)(68.00,8.50)(66.00,8.00)
\bezier{15}(54.00,12.00)(60.00,13.60)(66.00,12.00)
\bezier{5}(50.10,10.00)(52.00,11.50)(54.00,12.00)
\bezier{5}(69.90,10.00)(68.00,11.50)(66.00,12.00)
\put(50.00,27.00){\line(1,0){2.00}}
\put(52.00,27.00){\line(0,-1){4.00}}
\put(50.00,23.00){\line(1,0){2.00}}
\put(70.00,27.00){\line(-1,0){2.00}}
\put(68.00,27.00){\line(0,-1){4.00}}
\put(70.00,23.00){\line(-1,0){2.00}}
\bezier{15}(50.10,36.00)(52.00,34.50)(54.00,34.00)
\bezier{60}(54.00,34.00)(60.00,32.40)(65.60,33.80)
\bezier{15}(69.90,36.00)(68.00,34.50)(66.10,34.05)
\put(54.00,38.00){\vector(0,-1){24.50}}
\put(54.00,12.90){\line(0,-1){4.90}}
\bezier{100}(76.00,30.00)(89.00,32.00)(102.00,32.00)
\bezier{90}(76.00,15.00)(87.00,12.00)(98.00,12.00)
\put(100.00,17.50){\line(4,5){19.80}}
\put(150.00,17.50){\line(4,5){19.80}}
\put(100.00,7.50){\line(4,5){7.45}}
\put(150.00,7.50){\line(4,5){19.80}}
\put(120.00,42.50){\line(1,0){50.00}}
\put(170.00,32.50){\line(-1,0){7.45}}
\put(100.00,17.50){\line(1,0){50.00}}
\put(100.00,7.50){\line(1,0){50.00}}
\bezier{90}(124.60,34.20)(135.40,37.20)(147.40,34.00)
\bezier{90}(123.40,26.00)(135.40,22.40)(147.40,26.00)
\bezier{80}(122.40,34.00)(108.00,30.00)(123.40,26.00)
\bezier{80}(147.40,34.00)(162.80,30.00)(147.40,26.00)
\bezier{50}(145.20,42.50)(140.00,37.50)(137.00,30.00)
\bezier{10}(137.00,30.00)(134.00,24.00)(134.00,17.50)
\bezier{50}(134.00,17.50)(134.00,12.50)(135.00,9.10)
\thicklines
\put(20.75,33.95){\vector(2,1){0.30}}
\put(9.00,13.50){\vector(0,-1){0.30}}
\put(44.50,28.00){\vector(1,0){0.30}}
\put(65.75,33.95){\vector(2,1){0.30}}
\put(54.00,13.50){\vector(0,-1){0.30}}
\put(102.00,32.00){\vector(1,0){0.20}}
\put(98.00,12.00){\vector(1,0){0.20}}
\put(124.20,34.40){\vector(-4,-1){1.50}}
\put(135.00,9.10){\vector(1,-3){0.40}}
\put(50.00,25.00){\line(1,0){20.00}}
\put(129.60,30.00){\line(1,0){12.00}}
\bezier{10}(128.80,29.40)(129.40,30.00)(130.00,30.60)
\bezier{10}(140.80,29.40)(141.40,30.00)(142.00,30.60)
\end{picture}

\noindent
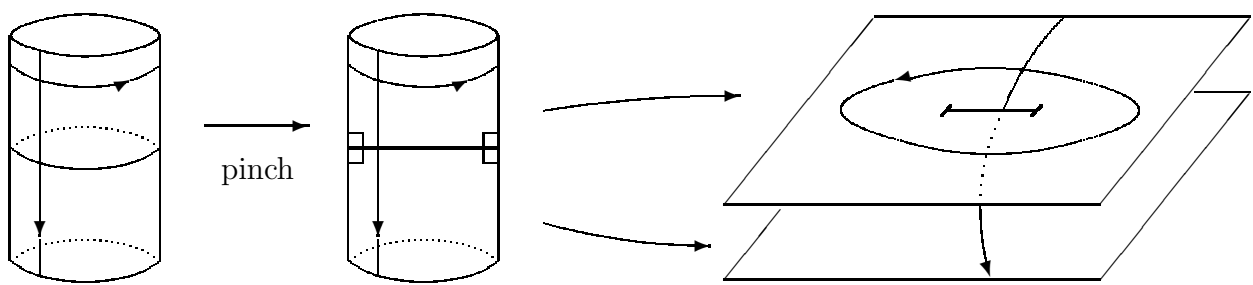
{\bf Fig.\ 2:} A cylinder is pinched into two bicones, which are mapped
to the two sheets of a Riemann surface. A $u =$ const. geodesic (circling
the cut) and a $v =$ const. geodesic (crossing the cut) are shown. 

%\end{document}
\newpage

%\documentstyle[bezier]{article}
%\textheight220mm\textwidth160mm\hoffset-15mm\voffset-25mm
%\begin{document}
%\unitlength 1.00mm
\begin{picture}(80.00,80.00)(-50,-20)
\bezier{100}(0.30,30.10)(5.25,40.50)(19.50,30.00)
\bezier{100}(0.00,28.50)(5.25,19.50)(19.50,30.00)
\bezier{45}(19.50,30.00)(23.25,33.30)(25.50,36.90)
\bezier{60}(26.50,38.50)(28.95,42.00)(30.00,48.00)
\bezier{45}(19.50,30.00)(23.25,26.70)(26.10,22.75)
\bezier{60}(27.00,21.00)(28.95,18.00)(30.00,12.00)
\bezier{12}(12.00,48.75)(15.45,42.75)(17.25,39.00)
\bezier{7}(17.50,38.00)(18.60,34.50)(19.50,30.00)
\bezier{12}(12.00,11.25)(15.45,17.25)(17.25,21.00)
\bezier{7}(17.50,22.00)(18.60,25.50)(19.50,30.00)
\bezier{9}(19.50,30.00)(24.00,30.00)(28.50,30.00)
\bezier{9}(30.00,30.00)(34.50,30.00)(39.00,30.00)
\put(12.00,52.00){\makebox(0,0)[cc]{$v=l$}}
\put(12.00,8.00){\makebox(0,0)[cc]{$v=-l$}}
\put(34.50,33.00){\makebox(0,0)[cc]{$v=l$}}
\put(34.50,27.00){\makebox(0,0)[cc]{$v=-l$}}
\put(30.00,8.00){\makebox(0,0)[cc]{$u=0$}}
\put(19.50,30.00){\circle*{1.50}}
\thicklines
\put(5.25,30.00){\line(1,0){4.50}}
\put(5.10,30.75){\line(0,-1){1.50}}
\put(9.90,30.75){\line(0,-1){1.50}}
\put(0.30,28.50){\vector(0,1){1.00}} 
\put(25.95,37.05){\vector(1,1){1.00}}
\put(27.15,21.15){\vector(-1,1){1.00}}
\put(16.50,42.00){\vector(1,-3){1.00}}
\put(16.50,18.00){\vector(1,3){1.00}}
\put(28.50,30.00){\vector(1,0){1.00}}
\end{picture}

\noindent
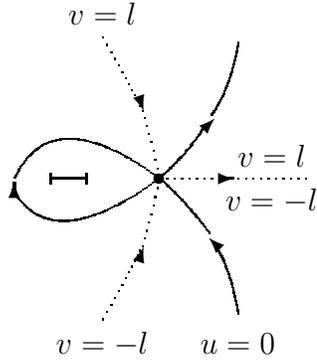
{\bf Fig.\ 3:} Some u-- and v--geodesics of the
dipole geometry; the critical geodesics $v = \pm l$ bissect the angles
formed by the self-intersecting critical geodesic $ u = 0$.
%\end{document}

%\documentstyle[bezier]{article}
%\textheight220mm\textwidth160mm\hoffset-15mm\voffset-25mm
%\begin{document}
%\unitlength 1.00mm
\begin{picture}(180.00,80.00)(80,-10)
\bezier{80}(95.60,10.00)(96.70,15.35)(110.00,16.00)
\bezier{80}(124.40,10.00)(123.30,15.35)(110.00,16.00)
\bezier{80}(95.60,10.00)(96.70,4.65)(110.00,4.00)
\bezier{80}(124.40,10.00)(123.30,4.65)(110.00,4.00)
\bezier{140}(83.90,10.00)(84.50,20.70)(110.00,22.00)
\bezier{140}(136.10,10.00)(135.50,20.70)(110.00,22.00)
\bezier{140}(83.90,10.00)(84.50,-0.30)(110.00,-2.00)
\bezier{140}(136.10,10.00)(135.50,-0.30)(110.00,-2.00)
\bezier{100}(91.75,10.00)(93.00,17.25)(109.00,18.10)
\bezier{100}(128.25,10.00)(127.00,17.25)(110.50,18.10)
\bezier{100}(91.75,10.00)(93.00,2.75)(109.50,1.90)
\bezier{100}(128.25,10.00)(127.00,2.75)(111.00,1.90)
\bezier{5}(124.40,10.00)(125.40,12.30)(130.25,13.00)
\bezier{5}(136.10,10.00)(135.10,12.30)(130.25,13.00)
\bezier{20}(124.40,10.00)(125.40,7.70)(130.25,7.00)
\bezier{20}(136.10,10.00)(135.10,7.70)(130.25,7.00)
\bezier{100}(181.75,10.00)(183.00,19.65)(199.00,20.80)
\bezier{100}(218.25,10.00)(217.00,19.65)(200.50,20.80)
\bezier{20}(181.75,10.00)(183.00,0.35)(199.50,-0.80)
\bezier{20}(218.25,10.00)(217.00,0.35)(201.00,-0.80)
\put(183.50,21.00){\makebox(0,0)[c,c]{$P$}}
\put(216.50,-1.00){\makebox(0,0)[c,c]{$P'$}}
\put(186.75,17.50){\circle*{1.00}}
\put(213.25,2.50){\circle*{1.00}}
\thicklines
\put(110.50,18.10){\vector(-1,0){1.20}}
\put(109.50,1.90){\vector(1,0){1.20}}
\put(83.90,10.00){\line(1,0){11.70}}
\put(200.50,20.80){\vector(-1,0){1.20}}
\put(199.50,-0.80){\vector(1,0){1.20}}
\put(175.90,10.00){\line(1,0){9.70}}
\put(214.40,10.00){\line(1,0){9.70}}
\put(175.90,9.00){\line(0,1){2.00}}
\put(185.60,9.00){\line(0,1){2.00}}
\put(214.40,9.00){\line(0,1){2.00}}
\put(224.10,9.00){\line(0,1){2.00}}
\end{picture}

\noindent
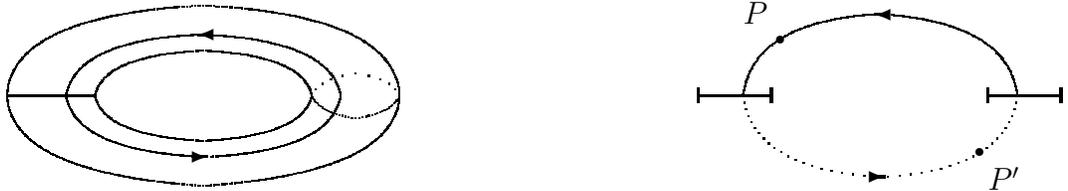
{\bf Fig.\ 4:} A torus is pinched into a tetracone, mapped to a
one-sheeted Riemann surface with two cuts identified under the involution
$P' \rightarrow P$.

%\end{document}
\newpage
%\documentstyle[bezier]{article}
%\textheight220mm\textwidth160mm\hoffset-15mm\voffset-25mm
%\begin{document}
%\unitlength 1.00mm
\begin{picture}(80.00,80.00)(-5,-15)
\bezier{50}(0.15,30.10)(2.62,40.50)(9.75,30.00)
\bezier{50}(0.00,28.50)(2.62,19.50)(9.75,30.00)
\bezier{30}(9.75,30.00)(11.62,33.30)(12.75,36.90)
\bezier{45}(13.25,38.50)(14.45,42.00)(15.00,48.00)
\bezier{30}(9.75,30.00)(11.62,26.70)(13.05,22.75)
\bezier{45}(13.50,21.00)(14.45,18.00)(15.00,12.00)
\bezier{8}(6.00,48.75)(7.75,42.75)(8.65,39.00)
\bezier{5}(8.75,38.00)(9.30,34.50)(9.75,30.00)
\bezier{8}(6.00,11.25)(7.75,17.25)(8.65,21.00)
\bezier{5}(8.75,22.00)(9.30,25.50)(9.75,30.00)
\bezier{14}(9.75,30.00)(16.50,30.00)(23.25,30.00)
\bezier{14}(24.25,30.00)(31.00,30.00)(37.75,30.00)
\put(2.00,52.00){\makebox(0,0)[cc]{\small $v=l$}}
\put(2.00,8.00){\makebox(0,0)[cc]{\small $v=-l$}}
\put(16.00,8.00){\makebox(0,0)[cc]{\small $u=-d$}}
\bezier{50}(47.35,30.10)(44.88,40.50)(37.75,30.00)
\bezier{50}(47.50,28.50)(44.88,19.50)(37.75,30.00)
\bezier{30}(37.75,30.00)(35.88,33.30)(34.75,36.90)
\bezier{45}(34.25,38.50)(33.05,42.00)(32.50,48.00)
\bezier{30}(37.75,30.00)(35.88,26.70)(34.45,22.75)
\bezier{45}(34.00,21.00)(33.05,18.00)(32.50,12.00)
\bezier{8}(41.50,48.75)(39.75,42.75)(38.85,39.00)
\bezier{5}(38.75,38.00)(38.20,34.50)(37.75,30.00)
\bezier{8}(41.50,11.25)(39.75,17.25)(38.85,21.00)
\bezier{5}(39.00,22.00)(38.45,25.50)(38.00,30.00)
\put(45.50,52.00){\makebox(0,0)[cc]{\small $v=l$}}
\put(45.50,8.00){\makebox(0,0)[cc]{\small $v=-l$}}
\put(31.50,8.00){\makebox(0,0)[cc]{\small $u=d$}}
\thicklines
\put(44.40,30.00){\line(-1,0){2.20}}
\put(44.70,30.75){\line(0,-1){1.50}}
\put(42.50,30.75){\line(0,-1){1.50}}
\put(3.10,30.00){\line(1,0){2.20}}
\put(2.80,30.75){\line(0,-1){1.50}}
\put(5.00,30.75){\line(0,-1){1.50}}
\put(9.75,30.00){\circle*{1.20}}
\put(37.75,30.00){\circle*{1.20}}
\put(0.15,28.50){\vector(0,1){1.00}}
\put(12.70,36.60){\vector(1,2){1.00}}
\put(14.00,20.70){\vector(-1,2){1.00}}
\put(7.80,42.00){\vector(1,-3){1.00}}
\put(8.25,18.00){\vector(1,3){1.00}}
\put(23.25,30.00){\vector(1,0){1.00}}
\put(47.35,28.50){\vector(0,1){1.00}}
\put(34.80,36.60){\vector(-1,2){1.00}}
\put(33.50,20.70){\vector(1,2){1.00}}
\put(38.50,38.00){\vector(1,3){1.00}}
\put(38.50,22.00){\vector(1,-3){1.00}}
\thinlines
\put(23.25,0.00){\makebox(0,0)[cc]{(a)}}
\put(95.00,45.00){\line(1,0){45.00}}
\put(80.00,30.00){\line(1,0){45.00}}
\put(102.50,40.00){\line(1,0){20.00}}
\put(97.50,35.00){\line(1,0){20.00}}
\put(80.00,30.00){\line(1,1){15.00}}
\put(125.00,30.00){\line(1,1){15.00}}
\put(97.50,35.00){\line(1,1){5.00}}
\put(100.00,35.00){\line(1,1){5.00}}
\put(102.50,35.00){\line(1,1){5.00}}
\put(105.00,35.00){\line(1,1){5.00}}
\put(107.50,35.00){\line(1,1){5.00}}
\put(110.00,35.00){\line(1,1){5.00}}
\put(112.50,35.00){\line(1,1){5.00}}
\put(115.00,35.00){\line(1,1){5.00}}
\put(117.50,35.00){\line(1,1){5.00}}

\bezier{20}(97.50,23.00)(98.80,25.60)(102.50,23.00)
\bezier{20}(97.50,23.00)(98.80,20.40)(102.50,23.00)
\bezier{20}(122.50,23.00)(121.20,25.60)(117.50,23.00)
\bezier{20}(122.50,23.00)(121.20,20.40)(117.50,23.00)
\bezier{50}(97.50,23.00)(98.50,13.00)(103.50,10.00)
\bezier{50}(122.50,23.00)(121.50,13.00)(116.50,10.00)
\bezier{50}(103.50,10.00)(110.00,5.00)(116.50,10.00)
\bezier{35}(102.50,23.00)(103.00,18.00)(105.00,15.50)
\bezier{35}(117.50,23.00)(117.00,18.00)(115.00,15.50)
\bezier{40}(105.00,15.50)(110.00,10.50)(115.00,15.50)
\bezier{15}(110.00,13.00)(112.00,10.50)(110.00,8.00)
\bezier{6}(110.00,13.00)(108.00,10.50)(110.00,8.00)
\thicklines
\put(117.00,41.00){\vector(-1,-1){1.00}}
\bezier{50}(116.50,41.00)(122.50,44.00)(126.00,40.00)
\bezier{35}(126.00,40.00)(128.00,37.50)(121.50,33.50)
\bezier{50}(121.50,33.50)(114.00,30.00)(112.00,34.00)
\put(112.00,34.00){\vector(-1,1){1.00}}
\put(101.50,35.50){\vector(-1,2){1.00}}
\bezier{20}(101.50,35.50)(103.00,31.00)(101.50,26.50)
\put(101.50,26.50){\vector(-1,-2){1.00}}
\put(121.50,35.50){\vector(-1,2){1.00}}
\bezier{20}(121.50,35.50)(123.00,31.00)(121.50,26.50)
\put(121.50,26.50){\vector(-1,-2){1.00}}
\put(102.50,40.00){\circle*{1.20}}
\put(122.50,40.00){\circle*{1.20}}
\put(97.50,35.00){\circle*{1.20}}
\put(117.50,35.00){\circle*{1.20}}
\put(102.50,23.00){\circle*{1.00}}
\put(117.50,23.00){\circle*{1.00}}
\thinlines
\put(112.50,42.50){\makebox(0,0)[cc]{$2d$}}
\put(97.50,38.50){\makebox(0,0)[cc]{$2l$}}
\put(110.00,5.00){\makebox(0,0)[cc]{$2L$}}
\bezier{10}(113.00,5.00)(114.50,5.50)(116.00,6.25)
\put(116.00,6.50){\vector(1,1){1.00}}
\bezier{10}(107.00,5.00)(105.50,5.50)(104.00,6.25)
\put(104.00,6.50){\vector(-1,1){1.00}}
\put(110.00,0.00){\makebox(0,0)[cc]{(b)}}
\end{picture}

\noindent
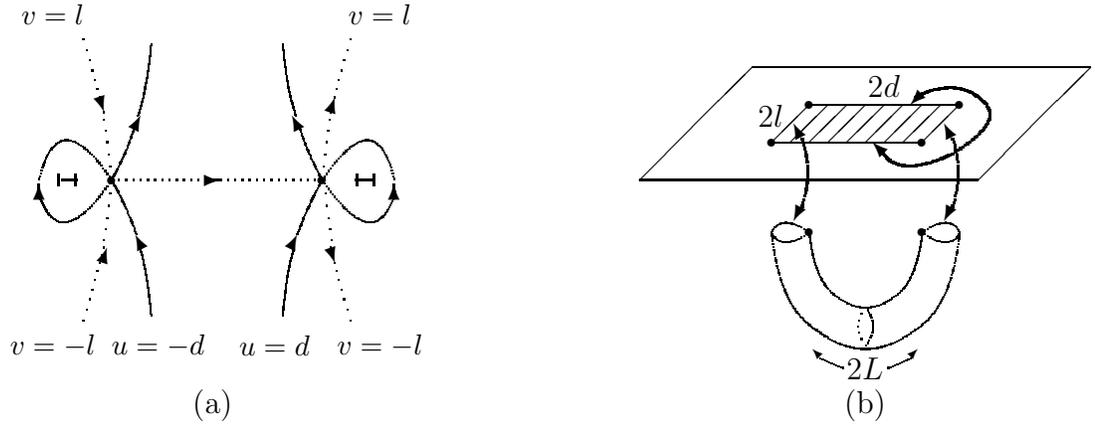
{\bf Fig.\ 5:} The DD Wheeler--Misner wormhole: (a) u-- and v--geodesics;
(b) geometrical construction.
%
%\end{document}

%\documentstyle[bezier]{article}
%\textheight220mm\textwidth160mm\hoffset-15mm\voffset-25mm
%\begin{document}
%\unitlength 1.00mm
\begin{picture}(80.00,80.00)(5,-5)
\bezier{50}(27.35,30.10)(24.88,36.00)(17.75,30.00)
\bezier{50}(27.50,28.50)(24.88,24.00)(17.75,30.00)
\bezier{30}(17.75,30.00)(14.50,33.30)(12.00,36.90)
\bezier{40}(11.00,38.50)(8.50,43.00)(7.30,48.00)
\bezier{30}(17.75,30.00)(14.50,26.70)(12.00,23.10)
\bezier{40}(11.00,21.50)(8.50,17.00)(7.30,12.00)
\put(7.30,8.00){\makebox(0,0)[cc]{\small $u=-d$}}
\bezier{6}(0.00,30.00)(4.00,30.00)(8.00,30.00)
\bezier{7}(9.00,30.00)(13.50,30.00)(17.75,30.00)
\bezier{50}(40.15,30.10)(42.62,36.00)(49.75,30.00)
\bezier{50}(40.00,28.50)(42.62,24.00)(49.75,30.00)
\bezier{30}(49.75,30.00)(53.00,33.30)(55.50,36.90)
\bezier{40}(56.50,38.50)(59.00,43.00)(60.20,48.00)
\bezier{30}(49.75,30.00)(53.00,26.70)(55.50,23.10)
\bezier{40}(56.50,21.50)(59.00,17.00)(60.20,12.00)
\put(60.20,8.00){\makebox(0,0)[cc]{\small $u=d$}}
\bezier{6}(67.50,30.00)(63.50,30.00)(59.50,30.00)
\bezier{7}(58.50,30.00)(54.00,30.00)(49.75,30.00)
\bezier{15}(17.75,30.00)(19.25,42.00)(33.25,43.50)
\bezier{15}(49.75,30.00)(48.25,42.00)(34.25,43.50)
\bezier{15}(17.75,30.00)(19.25,18.00)(33.25,16.50)
\bezier{15}(49.75,30.00)(48.25,18.00)(34.25,16.50)
\put(33.75,46.50){\makebox(0,0)[cc]{\small $v=l$}}
\put(33.75,13.50){\makebox(0,0)[cc]{\small $v=-l$}}
\put(33.75,0.00){\makebox(0,0)[cc]{(a)}}
\bezier{80}(103.75,30.50)(105.25,38.00)(119.25,39.00)
\bezier{80}(135.75,30.50)(134.25,38.00)(120.25,39.00)
\bezier{80}(103.75,29.50)(105.25,22.00)(119.25,21.00)
\bezier{80}(135.75,29.50)(134.25,22.00)(120.25,21.00)
\put(119.75,39.50){\line(0,1){5.00}}
\put(119.75,44.50){\line(0,1){3.50}}
\put(119.75,20.50){\line(0,-1){5.00}}
\put(119.75,14.50){\line(0,-1){3.50}}
\put(119.75,38.50){\line(0,-1){8.00}}
\put(119.75,29.50){\line(0,-1){8.00}}
\bezier{4}(113.50,30.50)(114.25,33.00)(115.00,34.25)
\bezier{4}(115.50,35.25)(117.75,37.60)(119.25,38.50)
\bezier{8}(120.25,39.50)(126.00,42.20)(131.75,43.75)
\bezier{6}(133.75,44.50)(138.25,45.60)(142.75,46.75)
\bezier{4}(126.00,30.50)(125.25,33.00)(124.50,34.25)
\bezier{4}(124.00,35.25)(121.75,37.60)(120.25,38.50)
\bezier{8}(119.25,39.50)(113.50,42.20)(107.75,43.75)
\bezier{6}(105.75,44.50)(101.25,45.60)(96.75,46.75)
\bezier{4}(113.50,29.50)(114.25,27.00)(115.00,25.75)
\bezier{4}(115.50,24.75)(117.75,22.40)(119.25,21.50)
\bezier{8}(120.25,20.50)(126.00,17.80)(131.75,16.25)
\bezier{6}(133.75,15.50)(138.25,14.40)(142.75,13.25)
\bezier{4}(126.00,29.50)(125.25,27.00)(124.50,25.75)
\bezier{4}(124.00,24.75)(121.75,22.40)(120.25,21.50)
\bezier{8}(119.25,20.50)(113.50,17.80)(107.75,16.25)
\bezier{6}(105.75,15.50)(101.25,14.40)(96.75,13.25)
\put(119.75,8.00){\makebox(0,0)[cc]{\small $u=0$}}
\put(89.00,13.25){\makebox(0,0)[cc]{\small $v=-d$}}
\put(89.00,46.25){\makebox(0,0)[cc]{\small $v=d$}}
\put(119.75,0.00){\makebox(0,0)[cc]{(b)}}
\put(17.75,30.00){\circle*{1.20}} 
\put(49.75,30.00){\circle*{1.20}}
\put(119.75,39.00){\circle*{1.20}} 
\put(119.75,21.00){\circle*{1.20}}
\thicklines
\put(24.40,30.00){\line(-1,0){2.20}}
\put(24.70,30.75){\line(0,-1){1.50}}
\put(22.50,30.75){\line(0,-1){1.50}}
\put(43.10,30.00){\line(1,0){2.20}}
\put(42.80,30.75){\line(0,-1){1.50}}
\put(45.00,30.75){\line(0,-1){1.50}}
\put(114.40,30.00){\line(-1,0){2.20}}
\put(114.50,30.75){\line(0,-1){1.50}}
\put(112.10,30.75){\line(0,-1){1.50}}
\put(125.10,30.00){\line(1,0){2.20}}
\put(125.20,30.75){\line(0,-1){1.50}}
\put(127.40,30.75){\line(0,-1){1.50}}
\put(27.35,28.50){\vector(0,1){1.50}}
\put(12.30,36.30){\vector(-1,2){1.50}}
\put(10.80,21.00){\vector(1,2){1.50}}
\put(8.00,30.00){\vector(1,0){1.00}}
\put(40.15,28.50){\vector(0,1){1.50}}
\put(55.20,36.30){\vector(1,2){1.50}}
\put(56.70,21.00){\vector(-1,2){1.50}}
\put(58.50,30.00){\vector(1,0){1.00}}
\put(33.25,43.50){\vector(1,0){1.00}}
\put(33.25,16.50){\vector(1,0){1.00}}
\put(103.75,29.50){\vector(0,1){1.50}}
\put(135.75,29.50){\vector(0,1){1.50}}
\put(119.75,43.50){\vector(0,1){1.50}}
\put(119.75,14.50){\vector(0,1){1.50}}
\put(119.75,30.50){\vector(0,-1){1.50}}
\put(116.00,35.25){\vector(-1,-1){1.00}}
\put(131.75,43.75){\vector(2,1){1.00}}
\put(123.50,35.25){\vector(-1,1){1.00}}
\put(107.75,43.75){\vector(2,-1){1.00}}
\put(116.00,24.75){\vector(-1,1){1.00}}
\put(131.75,16.25){\vector(2,-1){1.00}}
\put(123.50,24.75){\vector(-1,-1){1.00}}
\put(107.75,16.25){\vector(2,1){1.00}}
\end{picture}

\noindent
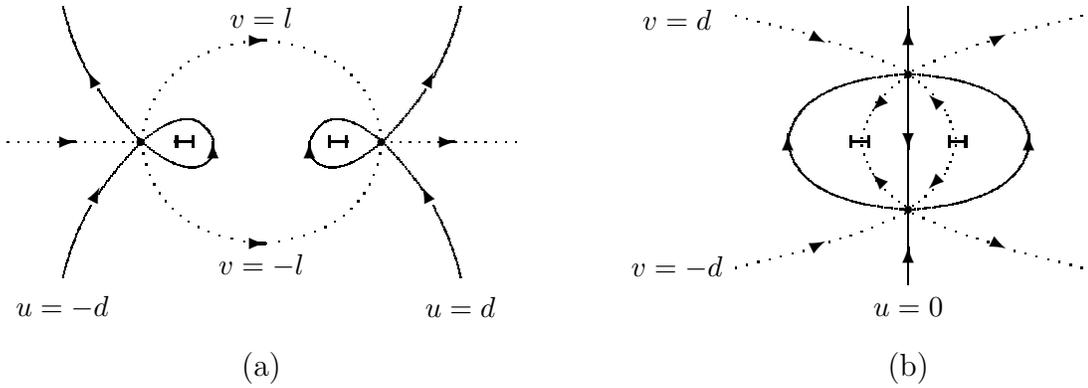
{\bf Fig.\ 6:} The comparison of the AA (a) and Q (b) geometries
leads to the equivalence of the corresponding Wheeler--Misner wormholes.
%
%\end{document}
\newpage
%\documentstyle[bezier]{article}
%\textheight220mm\textwidth160mm\hoffset-15mm\voffset-25mm
%\begin{document}
%\unitlength 1.00mm
\begin{picture}(80.00,50.00)(+65,-10)
\put(95.00,47.50){\line(1,0){45.00}}
\put(80.00,32.50){\line(1,0){45.00}}
\put(80.00,32.50){\line(1,1){15.00}}
\put(125.00,32.50){\line(1,1){15.00}}
\bezier{80}(95.60,10.00)(96.70,17.10)(110.00,18.00)
\bezier{80}(124.40,10.00)(123.30,17.10)(110.00,18.00)
\bezier{80}(95.60,10.00)(96.70,2.90)(110.00,2.00)
\bezier{80}(124.40,10.00)(123.30,2.90)(110.00,2.00)
\bezier{140}(83.90,10.00)(84.50,24.40)(110.00,26.00)
\bezier{140}(136.10,10.00)(135.50,24.40)(110.00,26.00)
\bezier{140}(83.90,10.00)(84.50,-4.40)(110.00,-6.00)
\bezier{140}(136.10,10.00)(135.50,-4.40)(110.00,-6.00)
\bezier{30}(110.00,-6.00)(104.00,-2.00)(110.00,2.00)
\bezier{8}(110.00,-6.00)(116.00,-2.00)(110.00,2.00)
\bezier{8}(110.00,26.00)(116.00,22.00)(110.00,18.00)
\bezier{30}(110.00,26.00)(104.00,22.00)(110.00,18.00)
\put(113.00,22.00){\circle*{1.20}}
\bezier{20}(89.75,10.00)(91.10,20.75)(110.00,22.00)
\bezier{20}(130.25,10.00)(129.15,20.75)(110.00,22.00)
\bezier{20}(89.75,10.00)(91.10,-0.75)(110.00,-2.00)
\bezier{20}(130.25,10.00)(129.15,-0.75)(110.00,-2.00)
\thicklines
\put(106.80,37.50){\line(1,0){1.60}}
\put(111.60,42.50){\line(1,0){1.60}}
\put(107.50,37.50){\line(1,1){5.00}}
\bezier{25}(107.80,20.35)(106.35,22.00)(107.70,24.30)
\put(107.00,20.00){\line(1,0){1.60}}
\put(107.00,24.00){\line(1,0){1.60}}
\bezier{50}(111.50,38.50)(115.00,34.00)(109.00,23.00)
\put(111.50,38.50){\vector(-1,2){0.80}}
\put(109.00,23.00){\vector(-1,-3){0.60}}
\thinlines
\put(115.50,44.00){\makebox(0,0)[cc]{$B$}}
\put(104.90,37.00){\makebox(0,0)[cc]{$A$}}
\put(105.00,19.50){\makebox(0,0)[cc]{$A'$}}
\put(106.00,27.50){\makebox(0,0)[cc]{$B'$}}
\put(116.00,22.00){\makebox(0,0)[cc]{$O$}}
\end{picture}

\noindent
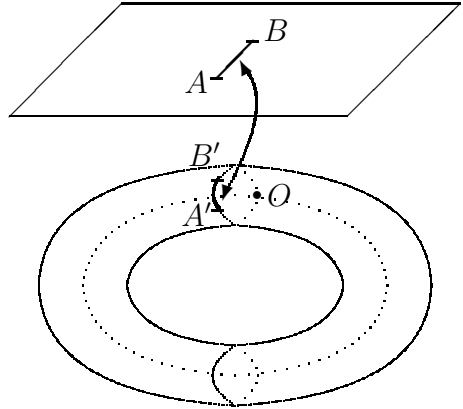
{\bf Fig.\ 7 :} Geometrical construction for the Q wormhole.
%\end{document}

%\documentstyle[bezier]{article}
%\textheight220mm\textwidth160mm\hoffset-15mm\voffset-25mm
%\begin{document}
%\unitlength 1.00mm
\begin{picture}(80.00,100.00)(10,0)
\put(25.00,80.00){\line(1,0){45.00}}
\put(10.00,65.00){\line(1,0){45.00}}
\put(10.00,65.00){\line(1,1){15.00}}
\put(55.00,65.00){\line(1,1){15.00}}
\put(75.00,72.50){\line(1,0){9.50}}
\put(105.00,80.00){\line(1,0){45.00}}
\put(90.00,65.00){\line(1,0){45.00}}
\put(90.00,65.00){\line(1,1){15.00}}
\put(135.00,65.00){\line(1,1){15.00}}
\bezier{40}(117.20,70.30)(114.50,74.00)(122.20,74.70)
\bezier{40}(117.80,70.30)(125.50,71.00)(122.80,74.70)
\bezier{10}(117.00,72.00)(118.00,73.00)(119.00,74.00)
\bezier{20}(117.30,70.50)(119.30,72.50)(121.30,74.50)
\bezier{10}(120.00,71.00)(121.50,72.50)(123.00,74.00)
\put(100.00,58.50){\line(-2,-1){14.50}}

\put(117.50,70.00){\circle*{1.20}}
\put(122.50,75.00){\circle*{1.20}}
\put(70.00,37.50){\circle*{1.20}}
\put(65.50,23.00){\circle*{1.20}}
\put(74.50,23.00){\circle*{1.20}}
\put(55.00,45.00){\line(1,0){45.00}}
\put(40.00,30.00){\line(1,0){45.00}}
\put(40.00,30.00){\line(1,1){15.00}}
\put(85.00,30.00){\line(1,1){15.00}}
\bezier{40}(70.20,37.70)(76.80,41.50)(78.00,37.50)
\bezier{40}(70.20,37.30)(76.80,33.50)(78.00,37.50)
\bezier{40}(69.80,37.70)(63.20,41.50)(62.00,37.50)
\bezier{40}(69.80,37.30)(63.20,33.50)(62.00,37.50)
\bezier{10}(62.50,37.00)(63.70,38.20)(64.90,39.40)
\bezier{15}(63.80,35.80)(65.30,37.30)(66.80,38.80)
\bezier{10}(66.60,36.10)(67.60,37.10)(68.60,38.10)
\bezier{10}(71.50,37.00)(72.50,38.00)(73.50,39.00)
\bezier{15}(73.40,36.20)(74.90,37.70)(76.40,39.20)
\bezier{10}(75.30,35.70)(76.60,37.00)(77.90,38.30)
\bezier{10}(58.00,22.50)(59.20,23.70)(60.40,24.90)
\bezier{15}(59.30,21.30)(60.80,22.80)(62.30,24.30)
\bezier{10}(62.10,21.60)(63.10,22.60)(64.10,23.60)
\bezier{10}(76.00,22.50)(77.00,23.50)(78.00,24.50)
\bezier{15}(77.90,21.70)(79.40,23.20)(80.90,24.70)
\bezier{10}(79.80,21.20)(81.00,22.40)(82.20,23.60)
\bezier{40}(57.50,23.00)(59.00,26.50)(65.30,23.00)
\bezier{40}(57.50,23.00)(59.00,19.50)(65.30,23.00)
\bezier{40}(82.50,23.00)(81.00,26.50)(74.70,23.00)
\bezier{40}(82.50,23.00)(81.00,19.50)(74.70,23.00)
\bezier{60}(57.50,23.00)(58.50,13.00)(63.50,10.00)
\bezier{60}(82.50,23.00)(81.50,13.00)(76.50,10.00)
\bezier{50}(63.50,10.00)(70.00,5.00)(76.50,10.00)
\bezier{30}(65.50,22.80)(65.90,18.00)(66.70,17.00)
\bezier{30}(74.50,22.80)(74.10,18.00)(73.30,17.00)
\bezier{40}(66.70,17.00)(70.00,11.50)(73.30,17.00)
\bezier{15}(70.00,14.00)(72.00,11.00)(70.00,8.00)
\bezier{6}(70.00,14.00)(68.00,11.00)(70.00,8.00)
\thicklines
\put(36.80,70.00){\line(1,0){1.60}}
\put(41.60,75.00){\line(1,0){1.60}}
\put(37.50,70.00){\line(1,1){5.00}}
\put(42.00,72.50){\vector(1,0){4.00}}
\put(38.00,72.50){\vector(-1,0){4.00}}
\put(113.50,65.50){\vector(1,1){3.00}}
\put(127.00,79.50){\vector(-1,-1){3.00}}
\put(84.50,72.50){\vector(1,0){0.50}}
\put(85.50,51.25){\vector(-2,-1){0.50}}
\put(63.50,32.40){\vector(1,2){1.00}}
\put(62.00,28.50){\vector(-1,-3){1.00}}
\bezier{50}(61.70,27.90)(62.30,31.00)(63.40,32.30)
\put(76.50,32.40){\vector(-1,2){1.00}}
\put(78.00,28.50){\vector(1,-3){1.00}}
\bezier{50}(78.30,27.90)(77.70,31.00)(76.60,32.30)
\end{picture}

\noindent 
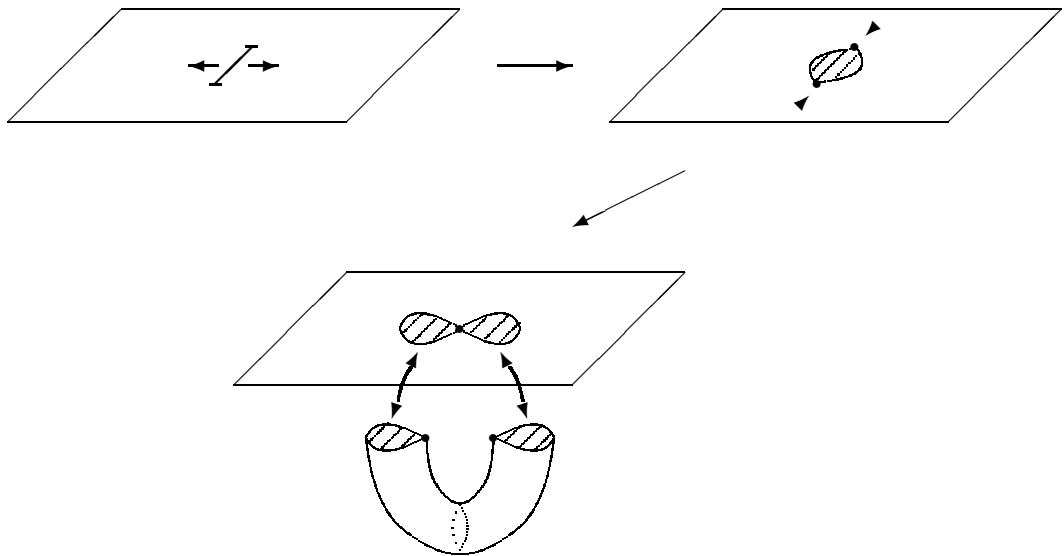
{\bf Fig.\ 8:} Geometrical construction for the 8 wormhole. 
%\end{document}
\newpage
%\documentstyle[bezier]{article}
%\textheight220mm\textwidth160mm\hoffset-15mm\voffset-25mm
%\begin{document}
\unitlength 1.3mm
\begin{picture}(80.00,40.00)(0,0)
\put(20.00,20.00){\line(1,0){20.00}}
\put(20.00,30.00){\line(1,0){20.00}}
\put(20.00,20.00){\line(0,1){10.00}}
\put(40.00,20.00){\line(0,1){10.00}}
\put(20.00,20.00){\circle*{1.20}}
\put(40.00,20.00){\circle*{1.20}}
\put(20.00,30.00){\circle*{1.20}}
\put(40.00,30.00){\circle*{1.20}}
\multiput(21,20)(2,0){5}{\line(1,1){10}}
\put(20,21){\line(1,1){9}}
\put(20,23){\line(1,1){7}}
\put(20,25){\line(1,1){5}}
\bezier{15}(20,27)(21.5,28.5)(23,30)
\put(31,20){\line(1,1){9}}
\put(33,20){\line(1,1){7}}
\put(35,20){\line(1,1){5}}
\bezier{15}(37,20)(38.5,21.5)(40,23)
\put(13,24){\line(1,0){7}}
\thicklines
\put(20.00,24.00){\line(-1,1){5.00}}
\put(14.50,29.50){\line(-1,1){4.00}}
\put(40.00,24.00){\line(1,-1){4.00}}
\put(44.50,19.50){\line(1,-1){4}}
\put(22.00,8.00){\line(-1,1){5.00}}
\put(16.50,13.50){\line(-1,1){10.00}}
\put(53.00,24.00){\line(-1,1){6.00}}
\put(46.50,30.50){\line(-1,1){9.00}}
\bezier{5}(27.00,30.00)(24.50,32.50)(22.00,35.00)
\bezier{4}(21.50,35.50)(19.50,37.50)(17.50,39.50)
\bezier{10}(27.00,20.00)(31.50,15.50)(36.00,11.00)
\put(14.50,29.50){\vector(1,-1){0.30}}
\put(16.50,13.50){\vector(1,-1){0.30}}
\put(44.00,20.00){\vector(1,-1){0.30}}
\put(21.50,35.50){\vector(1,-1){0.30}}
\put(46.50,30.50){\vector(1,-1){0.30}}
\put(36.00,11.00){\vector(1,-1){0.30}}
\thinlines
\put(8.5,36){\makebox(0,0)[cc]{1}}
\put(15.5,42.5){\makebox(0,0)[cc]{2}}
\put(4.5,26){\makebox(0,0)[cc]{3}}
\put(35,42){\makebox(0,0)[cc]{4}}
\put(13,27){\makebox(0,0)[cc]{$\theta$}}
\put(32,32.5){\makebox(0,0)[cc]{$2d$}}
\put(42,26){\makebox(0,0)[cc]{$2l$}}
\end{picture}

\noindent
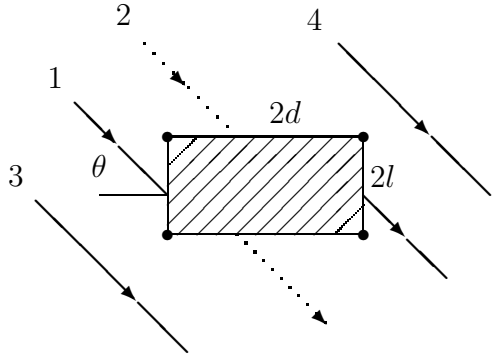
{\bf Fig.\ 9:} Geodesics through the DD$_0$ wormhole.

%\end{document}
%\documentstyle[bezier]{article}
%\textheight220mm\textwidth160mm\hoffset-15mm\voffset-25mm
%\begin{document}
%\bigskip
\unitlength 1.00mm
\begin{picture}(180,90)(10,0)
\put(20,10){\line(1,0){85}}
\put(20,30){\line(1,0){125}}
\put(60,50){\line(1,0){85}}
\put(20,10){\line(0,1){5}}
\multiput(60,5)(40,0){2}{\line(0,1){10}}
\put(20,25){\line(0,1){5}}
\multiput(60,25)(40,0){3}{\line(0,1){10}}
\multiput(60,45)(40,0){3}{\line(0,1){10}}
\thicklines
\put(5,20){\line(2,1){4}}
\put(5,20){\line(2,-1){4}}
\bezier{15}(7.6,18.7)(8.4,20)(7.6,21.3)
\multiput(20,15)(40,0){3}{\line(0,1){10}}
\multiput(60,35)(40,0){3}{\line(0,1){10}}
\multiput(19,15)(40,0){3}{\line(1,0){2}}
\multiput(19.2,25)(40,0){3}{\line(1,0){1.6}}
\multiput(59.2,35)(40,0){3}{\line(1,0){1.6}}
\multiput(59.2,45)(40,0){3}{\line(1,0){1.6}}
\multiput(45,25)(40,0){2}{\circle*{1.2}}
\multiput(85,45)(40,0){2}{\circle*{1.2}}
\bezier{40}(5,20)(45,32.5)(85,45)
\bezier{60}(5,20)(65,32.5)(125,45)
\bezier{40}(5,20)(45,22.5)(85,25)
\thinlines
\put(2,20){\makebox(0,0)[cc]{$O$}}
\put(17,26){\makebox(0,0)[cc]{$A$}}
\put(17,14){\makebox(0,0)[cc]{$B$}}
\put(48,25){\makebox(0,0)[cc]{$S$}}
\put(88,25){\makebox(0,0)[cc]{$S'$}}
\multiput(88,45)(40,0){2}{\makebox(0,0)[cc]{$S'$}}
\put(40,7){\makebox(0,0)[cc]{$2L$}}
\put(97,20){\makebox(0,0)[cc]{$2d$}}
\put(110,20){\makebox(0,0)[cc]{$2l$}}
\put(110,15){\vector(0,-1){4}}
\put(110,25){\vector(0,1){4}}
\end{picture}

\noindent
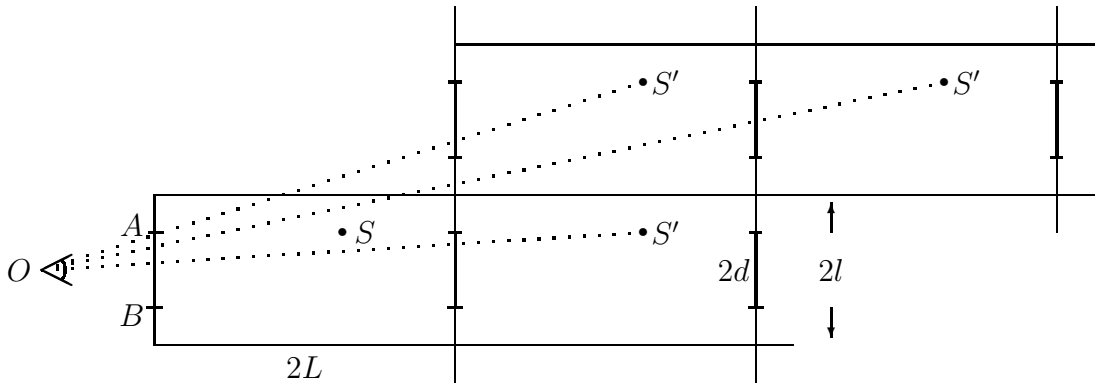
{\bf Fig.\ 10:} A Q wormhole gives a two--dimensional array of images $S'$
of a point source $S$.

\end{document}